# Integral representation for three-dimensional steady state size-dependent thermoelasticity

Ali R. Hadjesfandiari, Arezoo Hajesfandiari and Gary F. Dargush[*][†]
*Department of Mechanical and Aerospace Engineering*
*University at Buffalo, The State University of New York*
*Buffalo, NY 14260 USA*

April 6 2017

ABSTRACT

Boundary element methods provide powerful techniques for the analysis of problems involving coupled multi-physical response, especially in the linear case for which boundary-only formulations are possible. This paper presents the integral equation formulation for size-dependent linear thermoelastic response of solids under steady state conditions. The formulation is based upon consistent couple stress theory, which features a skew-symmetric couple-stress pseudo-tensor. For general anisotropic thermoelastic material, there is not only thermal strain deformation, but also thermal mean curvature deformation. Interestingly, in this size-dependent multi-physics model, the thermal governing equation is independent of the deformation. However, the mechanical governing equations depend on the temperature field. First, thermal and mechanical weak forms and reciprocal theorems are developed for this general size-dependent thermoelastic theory. Then, an integral equation formulation for the three-dimensional isotropic case is derived, along with the corresponding singular infinite space fundamental solutions or kernel functions. Remarkably, for isotropic materials within this theory, there is no thermal mean curvature deformation, and the thermoelastic effect is solely the result of thermal strain deformation. As a result, the size-dependent behavior is specified entirely by a single characteristic length scale parameter $l$, while the thermal coupling is defined in terms of the thermal expansion coefficient $\alpha$, as in the classical theory of steady state isotropic thermoelasticity. This simplification permits the development of the required kernel functions from previously defined fundamental solutions for isotropic media.

## 1. Introduction

As modern society becomes more technologically oriented and at the same time more energy conscious, there is a natural tendency to develop components and systems on progressively reduced length scales. At sufficiently small scales, the physical behavior of materials becomes size-dependent, which can affect the mechanical, thermal, electrical and magnetic performance of structures and mechanical components. Therefore, progress in micromechanics, nanomechanics and the associated technologies must go hand-in-hand with development of advanced size dependent models for coupled multi-physics phenomena, such as thermo- and electro-mechanics.

---

Of interest here is thermoelasticity, which is a multi-physics discipline concerned with predicting the linear thermo-mechanical behavior of solids subjected to both thermal and mechanical loadings. For modeling such phenomena near the smallest continuum scales, a size-dependent thermoelastic theory is needed.

Consistent size-dependent continuum mechanics theory can account for the length scale effect due to the microstructure of materials [1]. In this new theory, the second order couple-stress pseudo-tensor becomes skew-symmetric and is conjugate energetically to the mean curvature pseudo-tensor, which similarly is skew-symmetric. The resulting size-dependent continuum mechanics theory also provides a theoretical base to develop new size-dependent theories for a range of coupled multi-physics problems, including piezoelectricity [2] and thermoelasticity [3], where progress has already been made. Interestingly, in the corresponding size-dependent thermoelasticity [3], we find that the temperature can create thermal mean curvature deformation in addition to the usual thermal strain deformation. However, for isotropic materials, there is no thermal mean curvature deformation, and the thermoelastic effect is entirely the result of the thermal strain deformation. This means a temperature rise creates only strain deformation, and the appearance of couple-stresses is solely the result of the mechanical effect. Consequently, the size-dependency for isotropic materials is specified by one characteristic length scale parameter $l$, as in the isothermal elastic case. Meanwhile, the thermal coupling effect is specified through the thermal expansion coefficient $\alpha$, as in classical isotropic thermoelasticity.

Although the governing equations for the steady state size-dependent thermoelasticity are linear, analytical solutions are rare. In order for technology to take full advantage of this size-dependent phenomena, numerical methods for accurate modeling must be developed. This step is an important task in many micro- and nano-technological problems, where numerical formulations are needed to solve the general coupled boundary value problem. The boundary element method provides a suitable approach to solve the corresponding linear boundary value problems. This method is particularly attractive in cases where precision is required due to localized stress concentrations and stress intensities, or where one must deal with domains that extend toward infinity. The main difficulties in developing boundary element methods are the derivation of the corresponding integral equations and fundamental solutions, and the implementation of more complicated numerical techniques to evaluate the resulting singular and near singular integrals.

Boundary element methods for classical thermoelasticity have been developed long ago, for example, by Cruse et al. [4], Rizzo and Shippy [5], Tanaka et al. [6], Sladek and Sladek [7], Sharp and Crouch [8] and many other researchers [9, 10, 11]. Many articles have also appeared that employ boundary element methods to solve size-dependent elastic and thermoelastic problems within the framework of other non-classical theories, such as micropolar theory and second gradient elasticity. For example, boundary element methods in the former category have been developed by Das and Chaudhuri [12], Huang and Liang [13], Liang and Huang [14] and Shmoylova et al. [15]. On the other hand, second gradient elasticity boundary element formulations have been presented by Polyzos et al. [16], Tsepoura et al. [17] and Karlis et al. [18].

In the present work, we utilize the size-dependent thermoelastic theory from reference [3] and formulate the thermal and mechanical weak forms or virtual work theorems. Then, by using these as a base, we derive the corresponding thermal and mechanical reciprocal theorems for steady state size-dependent thermoelasticity for general linear anisotropic solids. These reciprocal theorems are, in turn, the foundation to formulate the corresponding integral equations. Within the current paper, the integral equations for isotropic thermoelastic materials are presented, along with the required singular infinite space Green's functions due to a point heat source, point force and point couple in the three-dimensional case. Interestingly, for the isotropic case, the size-dependent thermoelastic interactions simplify in such a way that all of these fundamental solutions can be constructed from previous work.

The balance of this paper is structured as follows. In Section 2, we provide an overview of steady state size-dependent thermoelasticity for general anisotropic materials. In Section 3, we develop for the first time the corresponding weak formulations or virtual work theorems. In this section, we also derive the new thermal and mechanical reciprocal theorems. Afterwards, we present the formulations for isotropic materials in Section 4. The reciprocal theorems are employed in Section 5 to derive the integral equations for three-dimensional steady state size-dependent thermoelastic response of isotropic media. This section also defines the corresponding three-dimensional kernel functions and includes a discussion of their singularities. Finally, we offer some conclusions in Section 6.

## 2. Consistent size-dependent steady state thermoelastic theory

In this section, we provide a brief overview of the size-dependent thermoelasticity theory for the general case of anisotropic materials. For a more detailed discussion on this theory, the reader is directed to reference [3].

Let us begin by considering an anisotropic elastic solid body occupying a volume $V$ enclosed by surface $S$. Assume that when this body is at a uniform absolute reference temperature $T_0$ and there is no external force applied, the body is undeformed and stress-free. Then, whenever the body is subjected to external forces and heat sources, it undergoes a temperature change (rise) field $\vartheta = T - T_0$, and an accompanying deformation specified by the displacement field $u_i$. These result in heat conduction and internal stresses in the body. The energy flow via heat conduction is represented by heat flux or thermal flux vector $q_i$. The scalar heat flux $q$ through a surface element $dS$ with unit outward normal vector $n_i$ is defined as

$$q = q_i n_i \tag{1}$$

The internal stresses are characterized by the force-stress tensor $\sigma_{ij}$, which is generally non-symmetric, and the couple-stress tensor $\mu_{ij}$. The force-traction vector $t_i$ and couple-traction vector $m_i$ acting on a surface element $dS$ with normal vector $n_i$ are given as

$$m_i = \mu_{ji} n_j \tag{2}$$
$$t_i = \sigma_{ji} n_j \tag{3}$$

In consistent couple stress theory [1], the couple-stress pseudo-tensor is skew-symmetric, that is

$$\mu_{ji} = -\mu_{ij} \tag{4}$$

The governing equations for steady state thermoelasticity are [3]:

Force equilibrium equation
$$\sigma_{ji,j} + F_i = 0 \tag{5}$$

Moment equilibrium equation
$$\mu_{ji,j} + \varepsilon_{ijk}\sigma_{jk} = 0 \tag{6}$$

Energy balance equation
$$-q_{i,i} + Q = 0 \tag{7}$$

where $F_i$ is the body force density, and $Q$ is the heat generated in unit time and unit volume, while $\varepsilon_{ijk}$ represents the permutation or Levi-Civita symbol.

Consistent continuum mechanics is founded on the idea that matter is continuously distributed in space. This requires the deformation to be specified entirely by the continuous displacement field $u_i$. Consequently, all kinematical quantities and deformation measures of must derive from this displacement field. Furthermore, consistent continuum mechanics theory should be an extension of rigid body mechanics, which then is recovered when there is no deformation. Thus, we must consider the rigid body portion of motion of infinitesimal elements of matter (or rigid triads) at each point of the continuum [19] and develop the kinematics from this perspective. This means that the displacement $u_i$ and rotation $\omega_i$ must be the kinematical degrees of freedom, since these are energetically conjugate to force-traction $t_i$ and couple-traction $m_i$, respectively. The infinitesimal strain and rotation tensors are defined as

$$e_{ij} = \frac{1}{2}\left(u_{i,j} + u_{j,i}\right) \tag{9}$$

$$\omega_{ij} = \frac{1}{2}\left(u_{i,j} - u_{j,i}\right) \tag{10}$$

respectively. A rotation vector $\omega_i$ also can be defined, as the dual to the rotation tensor, that is

$$\omega_i = \frac{1}{2}\varepsilon_{ijk}\omega_{kj} \tag{11}$$

The appropriate curvature measure for this theory is the mean curvature tensor, which can be written as the skew-symmetrical part of the rotation gradient, that is

$$\kappa_{ij} = \omega_{[i,j]} = \frac{1}{2}\left(\omega_{i,j} - \omega_{j,i}\right) \tag{12}$$

The true couple-stress vector $\mu_i$ and true mean curvature vector $\kappa_i$ dual to their corresponding tensors are defined as

$$\mu_i = \frac{1}{2}\varepsilon_{ijk}\mu_{kj} \tag{13}$$

$$\kappa_i = \frac{1}{2}\varepsilon_{ijk}\kappa_{kj} \tag{14}$$

Naturally, we also have the inverse relations

$$\varepsilon_{jik}\mu_k = \mu_{ij} \tag{15}$$

$$\varepsilon_{jik}\kappa_k = \kappa_{ij} \tag{16}$$

As with any second order tensor, the generally non-symmetric force-stress tensor $\sigma_{ij}$ can be decomposed into symmetric $\sigma_{(ij)}$ and skew-symmetric $\sigma_{[ij]}$ parts, such that

$$\sigma_{ij} = \sigma_{(ij)} + \sigma_{[ij]} \tag{17}$$

Then, angular equilibrium equation (6) can be written simply as

$$\sigma_{[ji]} = \frac{1}{2}\varepsilon_{ijk}\mu_{lk,l} = -\mu_{[i,j]} \tag{18}$$

which allows us to write the total force-stress tensor in the form

$$\sigma_{ji} = \sigma_{(ji)} + \frac{1}{2}\varepsilon_{ijk}\mu_{lk,l} = \sigma_{(ji)} - \mu_{[i,j]} \tag{19}$$

With this established, the linear equilibrium equation reduces to the following:

$$\left[\sigma_{(ji)} + \mu_{[j,i]}\right]_{,j} + F_i = 0 \tag{20}$$

For linear thermal response, we make use of the Duhamel generalization of the classical Fourier heat conduction law [20], for general anisotropic media, which is defined by

$$q_i = -k_{ij}T_{,j} \tag{21}$$

where $k_{ij}$ is the second order thermal conductivity tensor. Onsager [21] used non-equilibrium statistical mechanics to demonstrate that the conductivity tensor is symmetric, that is

$$k_{ij} = k_{ji} \tag{22}$$

Interestingly, this symmetric character of $k_{ij}$ can be established for continuum mechanics by invoking arguments from tensor analysis and linear algebra without postulating infinitesimal reversible processes [22].

In addition, we may write the following set of coupled constitutive relations for linear size-dependent thermoelasticity [3]

$$\sigma_{(ji)} = A_{ijkl}e_{kl} + C_{ijkl}\kappa_{kl} - a_{ij}\vartheta \tag{23}$$

$$\mu_{ji} = B_{ijkl}\kappa_{kl} + C_{klij}e_{kl} - b_{ij}\vartheta \tag{24}$$

Here the tensors $A_{ijkl}$, $B_{ijkl}$ and $C_{ijkl}$ represent the elastic coefficients, while the tensors $a_{ij}$ and $b_{ij}$ represent the thermoelastic properties of the material. The symmetric true tensor $a_{ij}$ is the coupling term between the temperature change $\vartheta$ and the symmetric part of force-stress tensor $\sigma_{(ji)}$ in (23), and the skew-symmetric pseudo tensor $b_{ij}$ is the size-dependent coupling term between the temperature and the couple-stress tensor $\mu_{ji}$ in (24).

Due to the tensorial nature of the field variables, the following symmetry (and skew-symmetry) relations must hold:

$$A_{ijkl} = A_{klij} = A_{jikl} \tag{25}$$

$$B_{ijkl} = B_{klij} = -B_{jikl} \tag{26}$$

$$C_{ijkl} = C_{jikl} = -C_{ijlk} \tag{27}$$

$$a_{ij} = a_{ji} \tag{28}$$

$$b_{ij} = -b_{ji} \tag{29}$$

Alternatively, these constitutive relations (23) and (24) can be rewritten as

$$\sigma_{(ji)} = A_{ijkl}\left(e_{kl} - \alpha_{kl}\vartheta\right) + C_{ijkl}\left(\kappa_{kl} - \beta_{kl}\vartheta\right) \tag{30}$$

$$\mu_{ji} = B_{ijkl}\left(\kappa_{kl} - \beta_{kl}\vartheta\right) + C_{klij}\left(e_{kl} - \alpha_{kl}\vartheta\right) \tag{31}$$

where

$$A_{ijkl}\alpha_{kl} + C_{ijkl}\beta_{kl} = a_{ij} \tag{32}$$

$$C_{klij}\alpha_{kl} + B_{ijkl}\beta_{kl} = b_{ij} \tag{33}$$

The symmetric tensor $\alpha_{ij}$ is the classical tensorial coefficient of thermal expansion, while $\beta_{ij}$ defines a new size-dependent skew-symmetric tensor that may be identified as the coefficient of thermal bending or flexion tensor. These two tensors have the following symmetry and skew-symmetry relations

$$\alpha_{ij} = \alpha_{ji} \tag{34}$$

$$\beta_{ij} = -\beta_{ji} \tag{35}$$

For the most general anisotropic case, the number of distinct components in $\alpha_{ij}$ and $\beta_{ij}$ are 6 and 3, respectively. We should emphasize that the quantities $\alpha_{ij}\vartheta$ and $\beta_{ij}\vartheta$ in (30) and (31) are the thermal strain and thermal mean curvature deformation tensors, respectively, created by the temperature change $\vartheta = T - T_0$. In other words, the temperature rise can create not only thermal strain deformation, but also thermal mean curvature deformation.

We note that the effect of thermal flexion moduli always appears in conjunction with couple-stresses. If the couple stress coefficients ($B_{ijkl} = 0$, $C_{ijkl} = 0$) are neglected, then all other size-dependent effects must be neglected as well ($\beta_{ij} = 0$).

Because of the skew symmetric character in $B_{ijkl}$, $C_{ijkl}$, $b_{ij}$ and $\beta_{ij}$, we can define their dual forms $B_{ij}$, $C_{ijk}$, $b_i$ and $\beta_i$, where

$$B_{ijkl} = \frac{1}{4}\varepsilon_{ijp}\varepsilon_{klq}B_{pq} \tag{36}$$

$$C_{ijkl} = \frac{1}{2}C_{ijm}\varepsilon_{mlk} \tag{37}$$

$$b_{ij} = \frac{1}{2}\varepsilon_{mji}b_m \tag{38}$$

$$\beta_{ij} = \frac{1}{2}\varepsilon_{mji}\beta_m \tag{39}$$

Therefore, the constitutive equations (30) and (31) can be expressed in terms of the mean curvature vector as

$$\sigma_{(ji)} = A_{ijkl}\left(e_{kl} - \alpha_{kl}\vartheta\right) + C_{ijk}\left(\kappa_k - \beta_k\vartheta\right) \tag{40}$$

$$\mu_i = -\frac{1}{2}B_{ij}\left(\kappa_j - \beta_j\vartheta\right) - \frac{1}{2}C_{kji}\left(e_{kj} - \alpha_{kj}\vartheta\right) \tag{41}$$

where

$$A_{ijkl}\alpha_{kl} + C_{ijk}\beta_k = a_{ij} \tag{42}$$

$$C_{kji}\alpha_{kj} + B_{ij}\beta_j = b_i \tag{43}$$

By bringing (21) into the energy balance equation (7), and constitutive relations (40) and (41) into the linear equilibrium equation (5), one may write the governing equations for uncoupled steady state size-dependent homogeneous thermoelasticity in the following form:

$$k_{ij}T_{,ij} + Q = 0 \tag{44}$$

$$\begin{aligned}A_{ijkl}u_{k,lj} + \frac{1}{4}C_{ijk}\left(u_{m,mjk} - \nabla^2 u_{k,j}\right) + \frac{1}{4}C_{kmi}\nabla^2 u_{k,m} - \frac{1}{4}C_{kmj}u_{k,mij} \\ + \frac{1}{16}B_{ik}\left(\nabla^2 u_{m,mk} - \nabla^2\nabla^2 u_k\right) - \frac{1}{16}B_{jk}\left(u_{m,mkij} - \nabla^2 u_{k,ij}\right) \\ - a_{ij}T_{,j} - \frac{1}{4}b_i T_{,jj} + \frac{1}{4}b_j T_{,ij} + F_i = 0\end{aligned} \tag{45}$$

Meanwhile, for the normal heat flux and normal tractions, we obtain

$$q = q_i n_i = -k_{ij}T_{,j}n_i \tag{46}$$

$$\begin{aligned}t_i &= \sigma_{ji}n_j \\ &= \left(\begin{array}{c}A_{ijkl}e_{kl} + C_{ijk}\kappa_k - a_{ij}\vartheta + \frac{1}{4}B_{im}\kappa_{m,j} - \frac{1}{4}B_{jm}\kappa_{m,i} \\ + \frac{1}{4}C_{kmi}e_{km,j} - \frac{1}{4}C_{kmj}e_{km,i} - \frac{1}{4}b_i\vartheta_{,j} + \frac{1}{4}b_j\vartheta_{,i}\end{array}\right)n_j\end{aligned} \tag{47}$$

$$m_i = \varepsilon_{ijk}\mu_k n_j = \varepsilon_{ijk}\left(-\frac{1}{2}B_{km}\kappa_m - \frac{1}{2}C_{mnk}e_{mn} + \frac{1}{2}b_k\vartheta\right)n_j \tag{48}$$

For boundary conditions, one may specify temperature change $\vartheta$ or normal heat flux $q$

$$\vartheta = \overline{\vartheta} \quad \text{on } S_\vartheta \tag{49a}$$

$$q = \overline{q} \quad \text{on } S_q \tag{49b}$$

displacements $u_i$ or force-tractions $t_i$

$$u_i = \overline{u}_i \quad \text{on } S_u \tag{50a}$$

$$t_i = \overline{t}_i \quad \text{on } S_t \tag{50b}$$

and tangential rotation $\omega_i$ or bending couple-traction $m_i$

$$\omega_i = \overline{\omega}_i \quad \text{on } S_\omega \tag{51a}$$

$$m_i = \overline{m}_i \quad \text{on } S_m \tag{51b}$$

Here $S_\vartheta$, $S_u$ and $S_\omega$ are the portions of the surface at which the essential boundary values for the temperature change $\vartheta$, the displacement vector $u_i$ and the rotation $\omega_i$ are prescribed, respectively. Furthermore, $S_q$, $S_t$ and $S_m$ are the portions of the surface at which the natural boundary values for the normal heat flux $q$, the force-traction vector $t_i$ and the couple-traction $m_i$ are specified, respectively. In order to construct a well-posed boundary value problem, we must have

$$S_\vartheta \cup S_q = S, \quad S_\vartheta \cap S_q = \varnothing \tag{52a,b}$$

$$S_u \cup S_t = S, \quad S_u \cap S_t = \varnothing \tag{53a,b}$$

$$S_\omega \cup S_m = S, \quad S_\omega \cap S_m = \varnothing \tag{54a,b}$$

Consequently, at each boundary point in a well-posed three-dimensional size-dependent thermoelasticity problem, a total of six independent boundary conditions must be specified. Then, six boundary quantities remain to be determined at each boundary point.

It should be noticed that in a steady state condition, the thermal heat conduction equation (44) is uncoupled from the mechanical equation (45). This means the thermal equation does not depend on the mechanical equation. Therefore, the thermal equation (44) can be solved independent of the mechanical equation (45). However, solving the mechanical equation (45) for $u_i$ requires the temperature field $\vartheta$. Consequently, equations (44) and (45) may be referred to as the uncoupled steady state size-dependent thermoelasticity equations.

## 3. Weak formulations

Weak formulations or virtual work theorems have important roles in many aspects of theoretical continuum mechanics and related computational methods. For example, these theorems can be used to construct variational and integral equation methods, as will be done here. Weak formulations also can be instrumental in establishing conditions of uniqueness. Therefore, we next present a weak formulation for steady state size-dependent thermoelasticity. Afterwards, starting with this weak form, we develop the reciprocal theorems for steady state thermoelastic response of size-dependent linear anisotropic materials.

### 3.1. Virtual work or weak form

Suppose there is an arbitrary differentiable temperature variation $\delta\vartheta = \delta T$ and displacement variation $\delta u_i$ in $V$, where the corresponding vectors of heat flux and angular rotation become

$$\delta q_i = -k\delta T_{,i} \tag{55}$$

$$\delta\omega_i = \frac{1}{2}\varepsilon_{ijk}\delta u_{k,j} \tag{56}$$

Let us begin by multiplying equation (7) by the virtual temperature $\delta T$ and then integrating over the volume, which produces

$$\int_V \left(-q_{i,i} + Q\right)\delta T dV = 0 \tag{57}$$

Next, using the relation

$$-q_{i,i}\delta T = \left(-q_i \delta T\right)_{,i} + q_i \delta T_{,i} \tag{58}$$

and the divergence theorem, we may rewrite (57) in the following form

$$-\int_V q_i \delta T_{,i} dV = -\int_S q_i \delta T n_i dS + \int_V Q\delta\vartheta dV \tag{59}$$

Then, using $q = q_i n_i$ and $\delta\vartheta = \delta T$, this becomes

$$-\int_V q_i \delta\vartheta_{,i} dV = -\int_S q\delta\vartheta dS + \int_V Q\delta\vartheta dV \tag{60}$$

which is the thermal virtual work formulation.

Now, let us construct the corresponding mechanical formulation. This can be obtained by multiplying (5) by the virtual displacement $\delta u_i$ and integrating over the volume, and also multiplying (6) by the energy conjugate virtual angular rotation $\delta\omega_i$ and integrating the result over the volume as well. Consequently, we have the following two separate equations

$$\int_V \left(\sigma_{ji,j} + F_i\right)\delta u_i dV = 0 \tag{61}$$

$$\int_V \left(\mu_{ji,j} + \varepsilon_{ijk}\sigma_{jk}\right)\delta\omega_i dV = 0 \tag{62}$$

However, we also have the relations
$$\sigma_{ji,j}\delta u_i = (\sigma_{ji}\delta u_i)_{,j} - \sigma_{ji}\delta u_{i,j} \tag{63}$$
$$\mu_{ji,j}\delta\omega_i + \varepsilon_{ijk}\sigma_{jk}\delta\omega_i = (\mu_{ji}\delta\omega_i)_{,j} - \mu_{ji}\delta\omega_{i,j} - \sigma_{jk}\delta\omega_{jk} \tag{64}$$

After using the divergence theorem, the relations (61) and (62) can be rewritten, respectively,
$$\int_V \sigma_{ji}\delta u_{i,j}dV = \int_S t_i\delta u_i dS + \int_V F_i\delta u_i dV \tag{65}$$
$$\int_V \mu_{ji}\delta\omega_{i,j}dV - \int_V \sigma_{ji}\delta\omega_{ij}dV = \int_S m_i\delta\omega_i dS \tag{66}$$

Then, by adding (65) and (66), we obtain the mechanical principle of virtual work as
$$\int_V \sigma_{(ji)}\delta e_{ij}dV + \int_V \mu_{ji}\delta\kappa_{ij}dV = \int_S t_i\delta u_i dS + \int_S m_i\delta\omega_i dS + \int_V F_i\delta u_i dV \tag{67}$$

We notice here that these thermal and mechanical virtual work formulations are uncoupled. This is the peculiar character of steady state thermoelasticity, in which the thermal equation is independent of the mechanical equation. This is in contrast with most other coupled problems, such as size-dependent piezoelectricity [2], and transient coupled thermoelasticity [10], where there is only a single coupled virtual work formulation.

*3.2. Reciprocal theorems*

We are now in position to derive the reciprocal theorems for steady state size-dependent linear anisotropic thermoelastic response under arbitrary applied thermal and mechanical loads. Consider the following distinct size-dependent solutions:
$$\left\{\vartheta^{(1)}, u_i^{(1)}, \omega_i^{(1)}, e_{ij}^{(1)}, \kappa_{ij}^{(1)}, q^{(1)}, t_i^{(1)}, m_i^{(1)}, Q^{(1)}, F_i^{(1)}\right\}$$
and
$$\left\{\vartheta^{(2)}, u_i^{(2)}, \omega_i^{(2)}, e_{ij}^{(2)}, \kappa_{ij}^{(2)}, q^{(2)}, t_i^{(2)}, m_i^{(2)}, Q^{(2)}, F_i^{(2)}\right\}$$
for the domain under consideration. We start by applying the thermal virtual work principle in the following two forms
$$-\int_V q_i^{(1)}\vartheta_{,i}^{(2)}dV = -\int_S q^{(1)}\vartheta^{(2)}dS + \int_V Q^{(1)}\vartheta^{(2)}dV \tag{68}$$
and
$$-\int_V q_i^{(2)}\vartheta_{,i}^{(1)}dV = -\int_S q^{(2)}\vartheta^{(1)}dS + \int_V Q^{(2)}\vartheta^{(1)}dV \tag{69}$$

However, the symmetry of the conductivity tensor (22) shows that the left hand of (68) and (69) are the same, that is
$$q_i^{(1)}\vartheta_{,i}^{(2)} = q_i^{(2)}\vartheta_{,i}^{(1)} \tag{70}$$

Therefore, by comparing (68) and (69), we obtain the general thermal reciprocal theorem for steady state linear heat conduction

$$-\int_S q^{(1)}\vartheta^{(2)}dS + \int_V Q^{(1)}\vartheta^{(2)}dV = -\int_S q^{(2)}\vartheta^{(1)}dS + \int_V Q^{(2)}\vartheta^{(1)}dV \tag{71}$$

Now let us apply the mechanical virtual work principle in the forms

$$\int_V \left[\sigma_{ji}^{(1)}e_{ij}^{(2)} + \mu_{ji}^{(1)}\kappa_{ij}^{(2)}\right]dV = \int_S t_i^{(1)}u_i^{(2)}dS + \int_S m_i^{(1)}\omega_i^{(2)}dS + \int_V F_i^{(1)}u_i^{(2)}dV \tag{72}$$

and

$$\int_V \left[\sigma_{ji}^{(2)}e_{ij}^{(1)} \mu_{ji}^{(2)}\kappa_{ij}^{(1)}\right]dV = \int_S t_i^{(2)}u_i^{(1)}dS + \int_S m_i^{(2)}\omega_i^{(1)}dS + \int_V F_i^{(2)}u_i^{(1)}dV \tag{73}$$

By employing the constitutive relations (23) and (24), we find

$$\begin{aligned}\sigma_{(ji)}^{(1)}e_{ij}^{(2)} + \mu_{ji}^{(1)}\kappa_{ij}^{(2)} &= A_{ijkl}e_{kl}^{(1)}e_{ij}^{(2)} + C_{ijkl}\kappa_{kl}^{(1)}e_{ij}^{(2)} - a_{ij}\vartheta^{(1)}e_{ij}^{(2)} \\ &+ B_{ijkl}\kappa_{kl}^{(1)}\kappa_{ij}^{(2)} + C_{klij}e_{kl}^{(1)}\kappa_{ij}^{(2)} - b_{ij}\vartheta^{(1)}\kappa_{ij}^{(2)}\end{aligned} \tag{74}$$

$$\begin{aligned}\sigma_{(ji)}^{(2)}e_{ij}^{(1)} + \mu_{ji}^{(2)}\kappa_{ij}^{(1)} &= A_{ijkl}e_{kl}^{(2)}e_{ij}^{(1)} + C_{ijkl}\kappa_{kl}^{(2)}e_{ij}^{(1)} - a_{ij}\vartheta^{(2)}e_{ij}^{(1)} \\ &+ B_{ijkl}\kappa_{kl}^{(2)}\kappa_{ij}^{(1)} + C_{klij}e_{kl}^{(2)}\kappa_{ij}^{(1)} - b_{ij}\vartheta^{(2)}\kappa_{ij}^{(1)}\end{aligned} \tag{75}$$

Symmetry relations (25)-(29) show that the left hand sides of (74) and (75) are the same and

$$\sigma_{ji}^{(1)}e_{ij}^{(2)} + \mu_{ji}^{(1)}\kappa_{ij}^{(2)} + \vartheta^{(1)}\left[a_{ij}e_{ij}^{(2)} + b_{ij}\kappa_{ij}^{(2)}\right] = \sigma_{ji}^{(2)}e_{ij}^{(1)} + \mu_{ji}^{(2)}\kappa_{ij}^{(1)} + \vartheta^{(2)}\left[a_{ij}e_{ij}^{(1)} + b_{ij}\kappa_{ij}^{(1)}\right] \tag{76}$$

Therefore, after making substitutions and comparing (72) and (73), we derive the mechanical reciprocal theorem for steady state size-dependent response of a linear anisotropic thermoelastic solid in the following form:

$$\begin{aligned}&\int_S \left[t_i^{(1)}u_i^{(2)} + m_i^{(1)}\omega_i^{(2)}\right]dS + \int_V F_i^{(1)}u_i^{(2)}dV + \int_V \left[a_{ij}e_{ij}^{(2)} + b_{ij}\kappa_{ij}^{(2)}\right]\vartheta^{(1)}dV \\ &= \int_S \left[t_i^{(2)}u_i^{(1)} + m_i^{(2)}\omega_i^{(1)}\right]dS + \int_V F_i^{(2)}u_i^{(1)}dV + \int_V \left[a_{ij}e_{ij}^{(1)} + b_{ij}\kappa_{ij}^{(1)}\right]\vartheta^{(2)}dV\end{aligned} \tag{77}$$

## 4. Formulation for linear isotropic thermoelastic materials

In this section, we restrict ourselves to the examination of size-dependent linear isotropic thermoelastic solids. For these materials, we have [3]

$$k_{ij} = k\delta_{ij} \tag{78}$$

$$A_{ijkl} = \lambda\delta_{ij}\delta_{kl} + \mu\delta_{ik}\delta_{jl} + \mu\delta_{il}\delta_{jk} \tag{79}$$

$$B_{ij} = 16\eta\delta_{ij} \tag{80}$$

$$C_{ijk} = 0 \tag{81}$$

$$\alpha_{ij} = \alpha\delta_{ij} \tag{82}$$

$$\beta_{ij} = 0, \quad \beta_i = 0 \tag{83}$$

$$a_{ij} = a\delta_{ij} \tag{84}$$

$$b_{ij} = 0, \quad b_i = 0 \tag{85}$$

where $\delta_{ij}$ is the Kronecker delta. The moduli $\lambda$ and $\mu$ are the Lamé constants for isotropic media in Cauchy elasticity, which are connected through the relationship

$$\lambda = 2\mu \frac{v}{1-2v} \tag{86}$$

with $v$ as Poisson's ratio. In addition, we find that

$$\lambda = \frac{vE}{(1+v)(1-2v)}, \quad \mu = \frac{E}{2(1+v)}, \quad 3\lambda + 2\mu = \frac{E}{1-2v} \tag{87}$$

where $E$ is Young's modulus of elasticity, while $K$ represents the bulk modulus of elasticity, with $K = \frac{E}{3(1-2v)}$.

For the isotropic solids, the coefficient of thermal flexion tensor $\beta_{ij}$ vanishes and there is only a single coefficient of thermal expansion $\alpha$, which implies that no thermal shear strain and thermal mean curvature will be created by the temperature change $\vartheta$. Consequently, we obtain for the coefficient $a$ the relations:

$$a = (3\lambda + 2\mu)\alpha = 3K\alpha = \frac{E}{1-2v}\alpha \tag{88}$$

The material constant $\eta$ relates the couple-stresses to mean curvatures in the isotropic material, where the ratio

$$\frac{\eta}{\mu} = l^2 \tag{89}$$

defines a characteristic material length $l$, which accounts for size-dependency in the infinitesimal deformation couple stress elasticity theory that we consider here.

We notice that for isotropic material $\beta = 0$, which shows that thermal mean curvature deformation will not develop. Therefore, for isotropic materials, the thermoelastic effect is solely the result of the thermal strain deformation. This means that a temperature rise only creates strain deformation, and any appearance of couple-stresses is the result of the mechanical effect.

As a result, the constitutive relations (21), (40) and (41) can be written

$$q_i = -k\vartheta_{,i} \tag{90}$$

$$\sigma_{(ji)} = \lambda e_{kk}\delta_{ij} + 2\mu e_{ij} - \frac{E}{1-2v}\alpha\vartheta\delta_{ij} \tag{91}$$

$$\mu_i = -8\mu l^2 \kappa_i \tag{92}$$

For the skew-symmetric part of the force-stress tensor in a homogeneous material, we have

$$\sigma_{[ji]} = -\mu_{[i,j]} = 2\mu l^2 \nabla^2 \omega_{ji}$$
$$= 2\mu l^2 \varepsilon_{ijk} \nabla^2 \omega_k \tag{93}$$

Then, combining (91) and (93), we may write

$$\sigma_{ji} = \lambda e_{kk} \delta_{ij} + 2\mu e_{ij} + 2\mu l^2 \nabla^2 \omega_{ji} - \frac{E}{1-2\nu}\alpha\vartheta\delta_{ij} \tag{94}$$

for the total force-stress tensor. Finally, the governing partial differential equations may be written:

$$k\nabla^2 T + Q = 0 \tag{95}$$

$$\left[\lambda + \mu\left(1 + l^2\nabla^2\right)\right] u_{k,ki} + \mu\left(1 - l^2\nabla^2\right)\nabla^2 u_i - \frac{E}{1-2\nu}\alpha T_{,i} + F_i = 0 \tag{96}$$

These are the governing partial differential equations for uncoupled steady state size-dependent linear thermoelasticity of a homogeneous isotropic solid.

For the normal heat flux and traction boundary values, we obtain

$$q = q_i n_i = -k T_{,i} n_i \quad \text{on } S \tag{97}$$

$$t_i = \sigma_{ji} n_j = \left[\lambda e_{kk}\delta_{ij} + 2\mu e_{ij} + 2\mu l^2 \nabla^2 \omega_{ji} - \frac{E}{1-2\nu}\alpha\vartheta\right] n_j \quad \text{on } S \tag{98}$$

$$m_i = \varepsilon_{ijk} n_j \mu_k = -8\varepsilon_{ijk}\mu l^2 \kappa_k n_j \quad \text{on } S \tag{99}$$

Consequently, at each boundary point in a well-posed three-dimensional size-dependent thermoelasticity problem, a total of six independent boundary conditions must be specified. The remaining six boundary quantities need to be determined as the boundary value problem solution.

Since for isotropic materials, $a_{ij} = a\delta_{ij}$, $b_{ij} = 0$, the thermal and mechanical reciprocal theorems for linear isotropic thermoelastic material reduce to

$$-\int_S q^{(1)}\vartheta^{(2)}dS + \int_V Q^{(1)}\vartheta^{(2)}dV = -\int_S q^{(2)}\vartheta^{(1)}dS + \int_V Q^{(2)}\vartheta^{(1)}dV \tag{100}$$

$$\int_S \left[t_i^{(1)}u_i^{(2)} + m_i^{(1)}\omega_i^{(2)}\right]dS + \int_V F_i^{(1)}u_i^{(2)}dV + \int_V ae_{kk}^{(2)}\vartheta^{(1)}dV$$
$$= \int_S \left[t_i^{(2)}u_i^{(1)} + m_i^{(2)}\omega_i^{(1)}\right]dS + \int_V F_i^{(2)}u_i^{(1)}dV + \int_V ae_{kk}^{(1)}\vartheta^{(2)}dV \tag{101}$$

We derive the boundary integral representation for the general three-dimensional isotropic case in the next section.

## 5. Integral representation for isotropic thermoelasticity

The thermal and mechanical reciprocal theorems (100) and (101) can be used to derive the boundary integral representation for the general three-dimensional steady state size-dependent isotropic thermoelasticity boundary value problem, which in turn can be used to formulate a boundary element method [23-27]. For this derivation, let one solution $^{(1)}$ in the reciprocal theorems represent the infinite space fundamental solutions due to a point heat source, point force and point couple, while solution $^{(2)}$ corresponds to the actual boundary value problem under consideration. To simplify the notation, we ignore the superscripts on this second solution and use $\{\vartheta, u_i, \omega_i, e_{ij}, q, t_i, m_i, Q, F_i\}$.

*5.1. Temperature integral equation*

In order to obtain the temperature integral equation, we first assume that in the infinitely extended thermoelastic isotropic body, a unit point heat source is applied at point $\xi$. For this first case, there is no body force. As a result, the loading is represented as

$$Q^{Q*} = \delta(\mathbf{x}-\xi), \quad F_i^{Q*} = 0 \tag{102}$$

where $\delta(\mathbf{x}-\xi)$ denotes the three-dimensional Dirac delta function. The heat source creates a temperature rise $\vartheta^{Q*}(\mathbf{x},\xi)$, which represents the temperature at $\mathbf{x}$ due to this unit heat point source at $\xi$. The governing equation for temperature change (rise) is the heat conduction equation

$$k\nabla^2 \vartheta^{Q*} + \delta(\mathbf{x},\xi) = 0 \tag{103}$$

which gives the weakly (*i.e.*, $\frac{1}{r}$) singular infinite space Green's function response or fundamental solution

$$\vartheta^{Q*}(\mathbf{x},\xi) = \frac{1}{4\pi k r} \tag{104}$$

It should be noticed that here the type of singular character of the kernels (weak, strong and hypersingular) is based on their behavior in surface integrals.

From (104), we obtain the flux field vector field as

$$q_i^{Q*}(\mathbf{x},\xi) = -k\vartheta_{,i}^{Q*} = \frac{1}{4\pi r^2} z_i \tag{105}$$

where

$$r^2 = (x_i - \xi_i)(x_i - \xi_i), \quad z_i = (x_i - \xi_i)/r \tag{106a,b}$$

Therefore, the normal heat flux field becomes

$$q^{Q*}(\mathbf{x},\xi) = q_i^{Q*} n_i = \frac{1}{4\pi r^2} z_i n_i \tag{107}$$

which contains a strong $1/r^2$ singularity.

The temperature rise $\vartheta^{Q*}(\mathbf{x},\boldsymbol{\xi})$ in turn creates some deformation $u_i^{Q*}(\mathbf{x},\boldsymbol{\xi})$ in the material, given by

$$u_i^{Q*}(\mathbf{x},\boldsymbol{\xi}) = \frac{\alpha}{8\pi k}\frac{1+\nu}{1-\nu}z_i \tag{108}$$

We notice that this bounded but non-analytic solution is identical to the solution in Cauchy or classical elasticity. Since this displacement is radial, its corresponding rotation is zero, that is

$$\omega_i^{Q*}(\mathbf{x},\boldsymbol{\xi}) = 0 \tag{109}$$

As a result, there is no mean curvature and couple stresses due to the point heat source in the isotropic body. Therefore, the couple-tractions kernel $m_i^{Q*}(\mathbf{x},\boldsymbol{\xi})$ vanishes, that is

$$m_i^{Q*}(\mathbf{x},\boldsymbol{\xi}) = 0 \tag{110}$$

The corresponding $\frac{1}{r}$ singular strain tensor due to the heat source is

$$e_{ij}^{Q*}(\mathbf{x},\boldsymbol{\xi}) = \frac{\alpha}{8\pi k}\frac{1+\nu}{1-\nu}\frac{1}{r}(\delta_{ij} - z_i z_j) \tag{111}$$

From this, we obtain

$$e_{kk}^{Q*}(\mathbf{x},\boldsymbol{\xi}) = \frac{\alpha}{4\pi k}\frac{1+\nu}{1-\nu}\frac{1}{r} \tag{112}$$

The constitutive relations show that the $\frac{1}{r}$ singular force-stress tensor due to the heat source is

$$\sigma_{ij}^{Q*}(\mathbf{x},\boldsymbol{\xi}) = -\mu\frac{\alpha}{4\pi k}\frac{1+\nu}{1-\nu}\frac{1}{r}(\delta_{ij} + z_i z_j) \tag{113}$$

Therefore, the corresponding weakly (i.e., $\frac{1}{r}$) singular force-traction kernel $t_i^{Q*}(\mathbf{x},\boldsymbol{\xi})$ can be written

$$t_i^{Q*}(\mathbf{x},\boldsymbol{\xi}) = -\mu\frac{\alpha}{4\pi k}\frac{1+\nu}{1-\nu}\frac{1}{r}(n_i + z_i z_j n_j) \tag{114}$$

Thus, we take the following definition of the infinite space Green's function as the first solution, which is due to a point heat source at $\boldsymbol{\xi}$:

$$\left\{\begin{array}{l}\vartheta^{Q^*}(\mathbf{x},\boldsymbol{\xi}), u_i^{Q^*}(\mathbf{x},\boldsymbol{\xi}), \omega_i^{Q^*}(\mathbf{x},\boldsymbol{\xi}), e_{ij}^{Q^*}(\mathbf{x},\boldsymbol{\xi}), q^{Q^*}(\mathbf{x},\boldsymbol{\xi}), t_i^{Q^*}(\mathbf{x},\boldsymbol{\xi}), \\ m_i^{Q^*}(\mathbf{x},\boldsymbol{\xi}), Q^{Q^*}(\mathbf{x},\boldsymbol{\xi}) = \delta(\mathbf{x}-\boldsymbol{\xi}), F_i^{Q^*}(\mathbf{x},\boldsymbol{\xi}) = 0\end{array}\right\} \quad (115)$$

Table 1 summarizes the singular behavior at point $\xi$ of the infinite space fundamental solutions in (115).

**Table 1**

Singularity of three-dimensional point heat source fundamental solutions.

| Size-dependent thermoelasticity kernel | Singularity |
|---|---|
| $\vartheta^{Q^*}(\mathbf{x},\boldsymbol{\xi})$ | weak singular $\dfrac{1}{r}$ |
| $u_i^{Q^*}(\mathbf{x},\boldsymbol{\xi})$ | regular |
| $\omega_i^{Q^*}(\mathbf{x},\boldsymbol{\xi}) = 0$ | - |
| $e_{ij}^{Q^*}(\mathbf{x},\boldsymbol{\xi})$ | singular $\dfrac{1}{r}$ |
| $q^{Q^*}(\mathbf{x},\boldsymbol{\xi})$ | strong singular $\dfrac{1}{r^2}$ |
| $t_i^{Q^*}(\mathbf{x},\boldsymbol{\xi})$ | weak singular $\dfrac{1}{r}$ |
| $m_i^{Q^*}(\mathbf{x},\boldsymbol{\xi}) = 0$ | - |

Using the relation (115) in the thermal reciprocal theorem (100), along with $\{\vartheta, u_i, \omega_i, e_{ij}, q, t_i, m_i, Q, F_i\}$, we obtain the following integral equation for the temperature at any point $\xi$ interior to the boundary $S$:

$$\vartheta(\boldsymbol{\xi}) - \int_S q^{Q^*}(\mathbf{x},\boldsymbol{\xi})\vartheta(\mathbf{x})dS(\mathbf{x}) = -\int_S \vartheta^{Q^*}(\mathbf{x},\boldsymbol{\xi})q(\mathbf{x})dS(\mathbf{x}) + \int_V \vartheta^{Q^*}(\mathbf{x},\boldsymbol{\xi})Q(\mathbf{x})dV(\mathbf{x}) \quad (116)$$

If we pass to the limit by letting $\xi$ approach $S$ from the inside, we may write

$$\begin{aligned}c^{Q^*}(\boldsymbol{\xi})\vartheta(\boldsymbol{\xi}) - \int_S q^{Q^*}(\mathbf{x},\boldsymbol{\xi})\vartheta(\mathbf{x})dS(\mathbf{x}) \\ = -\int_S \vartheta^{Q^*}(\mathbf{x},\boldsymbol{\xi})q(\mathbf{x})dS(\mathbf{x}) + \int_V \vartheta^{Q^*}(\mathbf{x},\boldsymbol{\xi})Q(\mathbf{x})dV(\mathbf{x})\end{aligned} \quad (117)$$

Here, the function $c^{Q*}(\xi)$ depends only on the local geometry at the boundary point $\xi$, which becomes equal to one-half for $\xi$ on a smooth portion of $S$. For non-smooth boundary points, $c^{Q*}(\xi)$ equals the interior solid angle in steradians divided by $4\pi$. Note that the integrands in (117) contain singularities as the integration point $\mathbf{x}$ approaches $\xi$. Although the integral involving $q^{Q*}(\mathbf{x},\xi)$ may appear to require treatment in the Cauchy principal value sense due to the strong singular $1/r^2$ character of the kernel, it turns out that for a boundary point $\xi$ this integral becomes regular. However, for the numerical implementation, we may use the constant temperature method to obtain the boundary element matrix coefficients associated with this integral. The other integrals in (117) are at most only weakly singular and may be evaluated directly with numerical quadrature.

Using (115) in the mechanical reciprocal theorem (101), along with $\{\vartheta, u_i, \omega_i, e_{ij}, q, t_i, m_i, Q, F_i\}$, we may also write the following integral equation

$$\int_S t_i^{Q*}(\mathbf{x},\xi) u_i(\mathbf{x}) dS(\mathbf{x}) = \int_S u_i^{Q*}(\mathbf{x},\xi) t_i(\mathbf{x}) dS(\mathbf{x}) + \int_V u_i^{Q*}(\mathbf{x},\xi) F_i(\mathbf{x}) dV(\mathbf{x}) + \int_V \alpha e_{kk}^{Q*}(\mathbf{x},\xi) \vartheta(\mathbf{x}) dV(\mathbf{x}) \quad (118)$$

Because there is no strong singular kernel in this integral equation, its form remains the same when $\xi$ is on the surface $S$. This weak singular integral equation is only one scalar equation, and will not be included in the final fundamental system of integral equations describing the steady state size-dependent thermoelasticity boundary value problem.

## 5.2. Displacement integral equation

Now we consider the infinite space Green's function response to the unit concentrated force at point $\xi$ in an arbitrary direction defined by the unit vector $e_i$. We also assume that there is no heat source density in this infinitely extended material. Therefore, the loading is represented as

$$Q^{F*} = 0, \quad F_i^{F*} = \delta(\mathbf{x}-\xi) e_i \quad (119)$$

As a result, there is no induced temperature field for the infinite domain in this case, that is

$$\vartheta^{F*} = 0, \quad q^{F*} = 0 \quad (120)$$

This shows that the thermal reciprocal theorem (100) is trivially satisfied without giving any equation.

Consequently, the infinite solution for the loading defined by (119) can be represented as

$$\left\{ \begin{array}{l} \vartheta^{F*}(\mathbf{x},\xi) = 0, u_i^{F*}(\mathbf{x},\xi), \omega_i^{F*}(\mathbf{x},\xi), e_{ik}^{F*}(\mathbf{x},\xi), q^{F*}(\mathbf{x},\xi) = 0, t_i^{F*}(\mathbf{x},\xi), \\ m_i^{F*}(\mathbf{x},\xi), Q^{F*}(\mathbf{x},\xi) = 0, F_i^{F*}(\mathbf{x},\xi) = \delta(\mathbf{x}-\xi) e_i \end{array} \right\} \quad (121)$$

where

$$\vartheta^{F*} = \vartheta_j^{F*} e_j \tag{122a}$$

$$u_i^{F*} = u_{ij}^{F*} e_j \tag{122b}$$

$$\omega_i^{F*} = \omega_{ij}^{F*} e_j \tag{122c}$$

$$e_{ik}^{F*} = e_{ikj}^{F*} e_j \tag{122d}$$

$$q^{F*} = q_j^{F*} e_j \tag{122e}$$

$$t_i^{F*} = t_{ij}^{F*} e_j \tag{122f}$$

$$m_i^{F*} = m_{ij}^{F*} e_j \tag{122g}$$

We notice that $u_{ij}^{F*}(\mathbf{x},\boldsymbol{\xi})$, $\omega_{ij}^{F*}(\mathbf{x},\boldsymbol{\xi})$, $t_{ij}^{F*}(\mathbf{x},\boldsymbol{\xi})$ and $m_{ij}^{F*}(\mathbf{x},\boldsymbol{\xi})$ are the displacement, rotation, force-traction and couple-traction, respectively, in the $i$-direction at $\mathbf{x}$ due to a unit point force in the $j$-direction at $\boldsymbol{\xi}$. Meanwhile, $e_{ikj}^{F*}(\mathbf{x},\boldsymbol{\xi})$ represents the corresponding strain tensor to the $i$- and $k$-directions at $\mathbf{x}$. Here, we have also included $\vartheta_j^{F*}(\mathbf{x},\boldsymbol{\xi})$ and $q_j^{F*}(\mathbf{x},\boldsymbol{\xi})$ for completeness, which are the temperature and normal heat flux response due to the unit point force in the $j$-direction at $\boldsymbol{\xi}$. Thus, we consider the infinite space fundamental solutions or Green's function to be the response to the point force in the $j$-direction at $\boldsymbol{\xi}$, that is

$$\left\{\begin{array}{l} \vartheta_j^{F*}(\mathbf{x},\boldsymbol{\xi}) = 0, u_{ij}^{F*}(\mathbf{x},\boldsymbol{\xi}), \omega_{ij}^{F*}(\mathbf{x},\boldsymbol{\xi}), e_{ikj}^{F*}(\mathbf{x},\boldsymbol{\xi}), q_j^{F*}(\mathbf{x},\boldsymbol{\xi}) = 0, t_{ij}^{F*}(\mathbf{x},\boldsymbol{\xi}), \\ m_{ij}^{F*}(\mathbf{x},\boldsymbol{\xi}), Q_j^{F*}(\mathbf{x},\boldsymbol{\xi}) = 0, F_{ij}^{F*}(\mathbf{x},\boldsymbol{\xi}) = \delta_{ij}\delta(\mathbf{x}-\boldsymbol{\xi}) \end{array}\right\} \tag{123}$$

The complete derivations of the elastic part of these infinite space Green's functions are given in reference [28].

The displacement fundamental solution can be written as

$$u_{ij}^{F*}(\mathbf{x},\boldsymbol{\xi}) = \frac{1}{16\pi\mu(1-\nu)}\frac{1}{r}\left[(3-4\nu)\delta_{iq} + z_i z_j\right]$$

$$+ \frac{1}{4\pi\mu}\frac{l^2}{r^3}\left[\left\{\left(3+3\frac{r}{l}+\frac{r^2}{l^2}\right)e^{-r/l} - 3\right\}z_i z_j + \left\{1-\left(1+\frac{r}{l}+\frac{r^2}{l^2}\right)e^{-r/l}\right\}\delta_{ij}\right] \tag{124}$$

The first term in Equation (124) is identical to the classical Cauchy elastic displacement kernel. We notice that $u_{ij}^{F*}(\mathbf{x},\boldsymbol{\xi})$ is weakly (i.e., $\frac{1}{r}$) singular.

The corresponding rotation vector due to the point force are

$$\omega_{ij}^{F*}(\mathbf{x},\boldsymbol{\xi}) = \frac{1}{8\pi\mu}\frac{1}{r^2}\left[1-\left(1+\frac{r}{l}\right)e^{-r/l}\right]\varepsilon_{ijk}z_k \tag{125}$$

By expanding the exponential factor, we find that $\omega_{ij}^{F*}(\mathbf{x},\boldsymbol{\xi})$ has no singularity.

The corresponding $\frac{1}{r^2}$ singular strain kernel tensor due to the point force is

$$e_{ikj}^{F*}(\mathbf{x},\boldsymbol{\xi}) =$$

$$= \frac{1}{32\pi\mu(1-v)}\frac{1}{r^2}\left[-(3-4v)\left(z_k\delta_{ij}+z_i\delta_{kj}\right)+2\delta_{ik}z_j+\delta_{kj}z_i+z_k\delta_{ij}-6z_iz_jz_k\right]a_q$$

$$+\frac{1}{4\pi\mu}\frac{l^2}{r^4}\left[\begin{array}{l}\left\{15-\left(15+15\frac{r}{l}+6\frac{r^2}{l^2}+\frac{r^3}{l^3}\right)e^{-r/l}\right\}z_iz_jz_k \\ +\left\{\left(3+3\frac{r}{l}+\frac{r^2}{l^2}\right)e^{-r/l}-3\right\}\left(\delta_{ik}z_j+\delta_{kj}z_i+\delta_{ij}z_k\right)\end{array}\right] \tag{126}$$

$$+\frac{1}{8\pi\mu}\frac{1}{r^2}\left(1+\frac{r}{l}\right)e^{-r/l}\left(z_i\delta_{kj}+z_k\delta_{ij}\right)$$

From this, we obtain

$$e_{kkj}^{F*} = -\frac{1-2v}{8\pi\mu(1-v)}\frac{z_j}{r^2} \tag{127}$$

For the corresponding force-tractions on a surface at $\mathbf{x}$ with outer normal $n_i$, we may write

$$t_{ij}^{F*}(\mathbf{x},\boldsymbol{\xi}) = -\frac{1}{8\pi(1-v)}\frac{1}{r^2}\left[(1-2v)\left(z_kn_k\delta_{ij}+z_in_j-z_jn_i\right)+3z_iz_jz_kn_k\right]$$

$$+\frac{1}{2\pi}\frac{l^2}{r^4}\left[\begin{array}{l}\left\{15-\left(15+15\frac{r}{l}+6\frac{r^2}{l^2}+\frac{r^3}{l^3}\right)e^{-r/l}\right\}z_iz_jz_kn_k \\ +\left\{\left(3+3\frac{r}{l}+\frac{r^2}{l^2}\right)e^{-r/l}-3\right\}\left(z_in_j+z_jn_i+\delta_{ij}z_kn_k\right)\end{array}\right] \tag{128}$$

$$+\frac{1}{2\pi}\frac{1}{r^2}\left[\left(1+\frac{r}{l}\right)e^{-r/l}\right]z_in_j$$

where the initial term in square brackets is the usual classical Cauchy elasticity traction kernel. We notice that $t_{ij}^{F*}(\mathbf{x},\boldsymbol{\xi})$ contains a strong $1/r^2$ singularity.

Meanwhile, the couple-traction on the surface at $\mathbf{x}$ is

$$m_{ij}^{F*} = \frac{1}{2\pi} \frac{l^2}{r^3} \left[ 3 - \left( 3 + 3\frac{r}{l} + \frac{r^2}{l^2} \right) e^{-r/l} \right] \left( z_i n_p \varepsilon_{pjk} - z_p n_p \varepsilon_{ijk} \right) z_k$$
$$+ \frac{1}{\pi} \frac{l^2}{r^3} \left[ 1 - \left( 1 + \frac{r}{l} \right) e^{-r/l} \right] \varepsilon_{ijk} n_k \tag{129}$$

These terms are weakly (*i.e.*, $\frac{1}{r}$) singular.

Table 2 summarizes the singular behavior at point $\boldsymbol{\xi}$ of the infinite space fundamental solutions in (123).

**Table 2**

Singularity of three-dimensional concentrated force fundamental solutions.

| Size-dependent thermoelasticity kernel | Singularity |
|---|---|
| $\vartheta_j^{F*}(\mathbf{x},\boldsymbol{\xi}) = 0$ | - |
| $u_{ij}^{F*}(\mathbf{x},\boldsymbol{\xi})$ | weak singular $\frac{1}{r}$ |
| $\omega_{ij}^{F*}(\mathbf{x},\boldsymbol{\xi})$ | regular |
| $e_{ikj}^{F*}(\mathbf{x},\boldsymbol{\xi})$ | singular $\frac{1}{r^2}$ |
| $q_j^{Q*}(\mathbf{x},\boldsymbol{\xi}) = 0$ | - |
| $t_{ij}^{F*}(\mathbf{x},\boldsymbol{\xi})$ | strong singular $\frac{1}{r^2}$ |
| $m_{ij}^{F*}(\mathbf{x},\boldsymbol{\xi})$ | weak singular $\frac{1}{r}$ |
| $f_j^{F*}(\mathbf{x},\boldsymbol{\xi})$ | regular |
| $h_j^{F*}(\mathbf{x},\boldsymbol{\xi})$ | weak singular $\frac{1}{r}$ |

Now for solution $^{(1)}$ in the reciprocal theorem, we introduce the infinite space fundamental solutions given by (123), while for solution $^{(2)}$, we choose the actual boundary value problem being considered $\{\vartheta, u_i, \omega_i, e_{ij}, q, t_i, m_i, Q, F_i\}$. After making the substitution of these two solutions into the mechanical reciprocal theorem (101) and invoking the sifting property of the Dirac delta function, we obtain the following integral representation for displacements at any point $\xi$ inside the bounding surface $S$:

$$u_j(\xi) + \int_S t_{ij}^{F*}(\mathbf{x},\xi) u_i(\mathbf{x}) dS(\mathbf{x}) + \int_S m_{ij}^{F*}(\mathbf{x},\xi) \omega_i(\mathbf{x}) dS(\mathbf{x})$$
$$= \int_S u_{ij}^{F*}(\mathbf{x},\xi) t_i(\mathbf{x}) dS(\mathbf{x}) + \int_S \omega_{ij}^{F*}(\mathbf{x},\xi) m_i(\mathbf{x}) dS(\mathbf{x}) + \int_V u_{ij}^{F*}(\mathbf{x},\xi) F_i(\mathbf{x}) dV(\mathbf{x}) \quad (130)$$
$$+ \int_V a e_{kkj}^{F*}(\mathbf{x},\xi) \vartheta(\mathbf{x}) dV(\mathbf{x})$$

The domain integral involving $e_{kkj}^{F*}(\mathbf{x},\xi)$ can be transformed to the boundary integral as in the case of classical isotropic thermoelasticity. For this we have

$$\int_V a e_{kkj}^{F*}(\mathbf{x},\xi) \vartheta(\mathbf{x}) dV(\mathbf{x}) = -\int_S h_j^{F*}(x,\xi) \vartheta(x) dS(x) + \int_S f_j^{F*}(x,\xi) q(x) dS(x) \quad (131)$$
$$- \int_V f_j^{F*}(\mathbf{x},\xi) Q(\mathbf{x}) dV(\mathbf{x})$$

where

$$f_j^{F*}(\mathbf{x},\xi) = \frac{\alpha}{8\pi k} \frac{1+\nu}{1-\nu} z_j \quad (132)$$

$$h_j^{F*}(\mathbf{x},\xi) = \frac{\alpha}{8\pi} \frac{1+\nu}{1-\nu} \frac{1}{r} (z_j z_k - \delta_{jk}) n_k \quad (133)$$

We notice that while $h_j^{F*}(\mathbf{x},\xi)$ is weakly singular, $f_j^{F*}(\mathbf{x},\xi)$ is not singular. Therefore, the integral equation (130) can be rewritten as

$$u_j(\xi) + \int_S t_{ij}^{F*}(\mathbf{x},\xi) u_i(\mathbf{x}) dS(\mathbf{x}) + \int_S m_{ij}^{F*}(\mathbf{x},\xi) \omega_i(\mathbf{x}) dS(\mathbf{x}) + \int_S h_j^{F*}(x,\xi) \vartheta(x) dS(x)$$
$$= \int_S u_{ij}^{F*}(\mathbf{x},\xi) t_i(\mathbf{x}) dS(\mathbf{x}) + \int_S \omega_{ij}^{F*}(\mathbf{x},\xi) m_i(\mathbf{x}) dS(\mathbf{x}) + \int_V u_{ij}^{F*}(\mathbf{x},\xi) F_i(\mathbf{x}) dV(\mathbf{x}) \quad (134)$$
$$+ \int_S f_j^{F*}(x,\xi) q(x) dS(x) - \int_V f_j^{F*}(\mathbf{x},\xi) Q(\mathbf{x}) dV(\mathbf{x})$$

Letting $\xi$ approach the bounding surface $S$ from inside the body, several of the integrands in (134) become singular due to the behavior of the fundamental solutions. In particular, the integral involving $t_{ij}^{F*}(\mathbf{x},\xi)$ must be evaluated in the Cauchy principal value sense. Then, for $\xi$ on $S$, the displacement boundary integral equation becomes

$$c_{ij}^{F*}(\xi)u_i(\xi)+ \oint_S t_{ij}^{F*}(\mathbf{x},\xi)u_i(\mathbf{x})dS(\mathbf{x})+\int_S m_{ij}^{F*}(\mathbf{x},\xi)\omega_i(\mathbf{x})dS(\mathbf{x})+\int_S h_j^{F*}(x,\xi)\vartheta(x)dS(x)$$
$$=\int_S u_{ij}^{F*}(\mathbf{x},\xi)t_i(\mathbf{x})dS(\mathbf{x})+\int_S \omega_{ij}^{F*}(\mathbf{x},\xi)m_i(\mathbf{x})dS(\mathbf{x})+\int_V u_{ij}^{F*}(\mathbf{x},\xi)F_i(\mathbf{x})dV(\mathbf{x}) \quad (135)$$
$$+\int_S f_j^{F*}(x,\xi)q(x)dS(x)-\int_V f_j^{F*}(\mathbf{x},\xi)Q(\mathbf{x})dV(\mathbf{x})$$

where $c_{ij}^{F*}(\xi)$ depends upon the local geometry at $\xi$ and takes a value identical to that for Cauchy elasticity. In particular, just as in the classical theory, for $\xi$ on a smooth portion of $S$, $c_{ij}^{F*}(\xi)$ reduces to the value $\frac{1}{2}\delta_{ij}$. The notation $\oint_S$ indicates that one must take the Cauchy principal value of the corresponding integral on $S$.

*5.3. Rotation integral equation*

To complete the required integral representations, we need to write the relation for rotations. This can be obtained either by taking the curl of equation (134) with respect to $\xi$ or by letting the first solution in the reciprocal theorem equal the infinite space Green's function for a point couple. We choose the latter approach and use the fundamental solution for a unit concentrated couple at point $\xi$ in the direction $e_i$. At the same time, we assume zero heat source density everywhere in the infinite material domain. Consequently, the applied loading is defined as

$$Q^{C*}=0, \quad C_i^{C*}=\delta(\mathbf{x}-\xi)e_i \quad (136)$$

For this loading, again there is no induced temperature field for the infinite domain in this case, thus

$$\vartheta^{C*}=0, \quad q^{C*}=0 \quad (137)$$

which means that the thermal reciprocal theorem (100) is satisfied trivially without providing an equation.

As established previously in [1, 28], the result of the concentrated couple can be represented as a body force represented by

$$F_i^{C*}=\frac{1}{2}\varepsilon_{ikj}\delta_{,k}(\mathbf{x}-\xi)e_j \quad (138)$$

along with a vanishing surface contribution at infinity.

The infinite space solution for the point couple loading defined by (136) can be represented by

$$\left\{\begin{array}{l}\vartheta^{C*}(\mathbf{x},\xi)=0, u_i^{C*}(\mathbf{x},\xi), \omega_i^{C*}(\mathbf{x},\xi), e_{ik}^{C*}(\mathbf{x},\xi), \kappa_{ik}^{C*}(\mathbf{x},\xi), q^{C*}(\mathbf{x},\xi)=0, t_i^{C*}(\mathbf{x},\xi), \\ m_i^{C*}(\mathbf{x},\xi), Q^{C*}(\mathbf{x},\xi)=0, F_i^{C*}(\mathbf{x},\xi)=\frac{1}{2}\varepsilon_{ikj}\delta_{,k}(\mathbf{x}-\xi)e_j\end{array}\right\} \quad (139)$$

where

$$\vartheta^{C*} = \vartheta_j^{C*} e_j \tag{140a}$$

$$u_i^{C*} = u_{ij}^{C*} e_j \tag{140b}$$

$$\omega_i^{C*} = \omega_{ij}^{C*} e_j \tag{140c}$$

$$e_{ik}^{C*} = e_{ikj}^{C*} e_j \tag{140d}$$

$$q^{C*} = q_j^{C*} e_j \tag{140e}$$

$$t_i^{C*} = t_{ij}^{C*} e_j \tag{140f}$$

$$m_i^{C*} = m_{ij}^{C*} e_j \tag{140g}$$

These solutions $u_{ij}^{C*}(\mathbf{x},\boldsymbol{\xi})$, $\omega_{ij}^{C*}(\mathbf{x},\boldsymbol{\xi})$, $t_{ij}^{C*}(\mathbf{x},\boldsymbol{\xi})$ and $m_{ij}^{C*}(\mathbf{x},\boldsymbol{\xi})$ represent, respectively, the displacement, rotation, force-traction and couple-traction in the $i$-direction at $\mathbf{x}$ due to a unit point couple in the $j$-direction at $\boldsymbol{\xi}$. Additionally, $e_{ikj}^{C*}(\mathbf{x},\boldsymbol{\xi})$ represents the corresponding strain tensor in the $i$- and $k$-directions at $\mathbf{x}$. Once again, we have also included $\vartheta_j^{C*}(\mathbf{x},\boldsymbol{\xi})$ and $q_j^{C*}(\mathbf{x},\boldsymbol{\xi})$ for completeness. These define the null temperature and normal heat flux fields due to the unit point couple in the $j$-direction at $\boldsymbol{\xi}$.

Thus, we consider the infinite space fundamental solutions or Green's function be the response to the point couple in the $j$-direction at $\boldsymbol{\xi}$, that is

$$\left\{ \begin{array}{l} \vartheta_j^{C*}(\mathbf{x},\boldsymbol{\xi}) = 0, u_{ij}^{C*}(\mathbf{x},\boldsymbol{\xi}), \omega_{ij}^{C*}(\mathbf{x},\boldsymbol{\xi}), e_{ikj}^{C*}(\mathbf{x},\boldsymbol{\xi}), q_j^{C*}(\mathbf{x},\boldsymbol{\xi}) = 0, t_{ij}^{C*}(\mathbf{x},\boldsymbol{\xi}), \\ m_{ij}^{C*}(\mathbf{x},\boldsymbol{\xi}), Q_j^{C*}(\mathbf{x},\boldsymbol{\xi}) = 0, F_{ij}^{C*}(\mathbf{x},\boldsymbol{\xi}) = \frac{1}{2}\varepsilon_{ikj}\delta_{,k}(\mathbf{x}-\boldsymbol{\xi}) \end{array} \right\} \tag{141}$$

The complete derivations of the elastic part of these infinite space Green's functions are given in reference [28].

The non-singular displacement fundamental solution can be written as

$$u_{ij}^{C*} = \omega_{ij}^{F*}(\mathbf{x},\boldsymbol{\xi}) = \frac{1}{8\pi\mu}\frac{1}{r^2}\left[1-\left(1+\frac{r}{l}\right)e^{-r/l}\right]\varepsilon_{ijk}z_k \tag{142}$$

As a result, the corresponding weakly singular rotation solution becomes

$$\omega_{ij}^{C*}(\mathbf{x},\boldsymbol{\xi}) = \frac{1}{16\pi\mu}\frac{1}{r^3}\left[3-\left(3+3\frac{r}{l}+\frac{r^2}{l^2}\right)e^{-r/l}\right](z_i z_j - \delta_{ij}) - \frac{1}{8\pi\mu}\frac{1}{r^3}\left[\left(1+\frac{r}{l}\right)e^{-r/l} - 1\right]\delta_{ij} \tag{143}$$

The corresponding strain tensor due to a unit point couple is

$$e_{ikj}^{C*} = \frac{1}{16\pi\mu} \frac{1}{r^3} \left[ 3 - \left(3 + 3\frac{r}{l} + \frac{r^2}{l^2}\right) e^{-r/l} \right] \left( z_k \varepsilon_{ipj} + z_i \varepsilon_{kpj} \right) z_p \tag{144}$$

Therefore, we notice that for this case

$$e_{kkj}^{C*} = 0 \tag{145}$$

Additionally, the force-traction vector becomes

$$t_{ij}^{C*}(\mathbf{x},\boldsymbol{\xi}) = -\frac{1}{8\pi} \frac{3}{r^3} z_k z_p \varepsilon_{ijp} n_k$$

$$+ \frac{1}{8\pi} \frac{1}{r^3} \left[ 3 - 2\left(3 + 3\frac{r}{l} + \frac{r^2}{l^2}\right) e^{-r/l} \right] z_i z_p \varepsilon_{pjk} n_k + \frac{1}{4\pi} \frac{1}{r^3} \left(1 + \frac{r}{l}\right) e^{-r/l} \varepsilon_{ijk} n_k \tag{146}$$

while the couple-traction fundamental solution assumes the form

$$m_{ij}^{C*}(\mathbf{x},\boldsymbol{\xi}) = \frac{1}{4\pi} \frac{1}{r^2} \left[ \left(1 + \frac{r}{l}\right) e^{-r/l} \right] \left( z_i n_j - z_k n_k \delta_{ij} \right) \tag{147}$$

The force-tractions $t_{ij}^{C*}(\mathbf{x},\boldsymbol{\xi})$ in (146) are hypersingular (i.e., $1/r^3$), while the couple-traction $m_{ij}^{C*}(\mathbf{x},\boldsymbol{\xi})$ in (147) contains a strong $1/r^2$ singularity.

The singularity of the infinite space fundamental solutions or Green's functions in Equation (141) are summarized in Table 3.

Using these results in the mechanical reciprocal theorem (101), along with $\{\vartheta, u_i, \omega_i, e_{ij}, q, t_i, m_i, Q, F_i\}$, leads to the following integral representation for rotation at any point $\boldsymbol{\xi}$ interior to the bounding surface $S$:

$$\omega_j(\boldsymbol{\xi}) + \int_S t_{ij}^{C*}(\mathbf{x},\boldsymbol{\xi}) u_i(\mathbf{x}) dS(\mathbf{x}) + \int_S m_{ij}^{C*}(\mathbf{x},\boldsymbol{\xi}) \omega_i(\mathbf{x}) dS(\mathbf{x})$$

$$= \int_S u_{ij}^{C*}(\mathbf{x},\boldsymbol{\xi}) t_i(\mathbf{x}) dS(\mathbf{x}) + \int_S \omega_{ij}^{C*}(\mathbf{x},\boldsymbol{\xi}) m_i(\mathbf{x}) dS(\mathbf{x}) + \int_V u_{ij}^{C*}(\mathbf{x},\boldsymbol{\xi}) F_i(\mathbf{x}) dV(\mathbf{x}) \tag{148}$$

If we pass to the limit as $\boldsymbol{\xi}$ approaches the surface $S$ from inside the body, we may establish the following boundary integral representation:

$$c_{ij}^{C*}(\boldsymbol{\xi}) \omega_i(\boldsymbol{\xi}) + \oint_S t_{ij}^{C*}(\mathbf{x},\boldsymbol{\xi}) u_i(\mathbf{x}) dS(\mathbf{x}) + \oint_S m_{ij}^{C*}(\mathbf{x},\boldsymbol{\xi}) \omega_i(\mathbf{x}) dS(\mathbf{x})$$

$$= \int_S u_{ij}^{C*}(\mathbf{x},\boldsymbol{\xi}) t_i(\mathbf{x}) dS(\mathbf{x}) + \int_S \omega_{ij}^{C*}(\mathbf{x},\boldsymbol{\xi}) m_i(\mathbf{x}) dS(\mathbf{x}) + \int_V u_{ij}^{C*}(\mathbf{x},\boldsymbol{\xi}) F_i(\mathbf{x}) dV(\mathbf{x}) \tag{149}$$

Table 3

Singularity of three-dimensional concentrated couple fundamental solutions.

| Size-dependent thermoelasticity kernel | Singularity |
|---|---|
| $\vartheta_j^{C*}(\mathbf{x},\boldsymbol{\xi}) = 0$ | - |
| $u_{ij}^{C*}(\mathbf{x},\boldsymbol{\xi})$ | regular |
| $\omega_{ij}^{C*}(\mathbf{x},\boldsymbol{\xi})$ | weak singular $\dfrac{1}{r}$ |
| $e_{ikj}^{C*}(\mathbf{x},\boldsymbol{\xi})$ | singular $\dfrac{1}{r}$ |
| $q_j^{C*}(\mathbf{x},\boldsymbol{\xi}) = 0$ | - |
| $t_{ij}^{C*}(\mathbf{x},\boldsymbol{\xi})$ | hypersingular $\dfrac{1}{r^3}$ |
| $m_{ij}^{C*}(\mathbf{x},\boldsymbol{\xi})$ | strong singular $\dfrac{1}{r^2}$ |

where now the function $c_{ij}^{C*}(\boldsymbol{\xi})$ depends only upon the local geometry at $\boldsymbol{\xi}$. For $\boldsymbol{\xi}$ on a smooth portion of $S$, $c_{ij}^{C*}(\boldsymbol{\xi})$ assumes the value $\dfrac{1}{2}\delta_{ij}$. Note again that several of the integrands in Equation (149) are singular, including the kernel $m_{ij}^{C*}(\mathbf{x},\boldsymbol{\xi})$, which is strongly singular. As a result, its corresponding integral needs to be evaluated in the Cauchy principal value sense. On the other hand, the integral corresponding to the hypersingular kernel $t_{ij}^{C*}(\mathbf{x},\boldsymbol{\xi})$ needs to be treated as Hadamard finite part, as signified by the symbol $\fint_S$ in (149). Meanwhile, the other integrals in (149) are at most weakly singular.

The relation expressed in equation (149) completes the definition of the boundary integral representation for three-dimensional steady state size-dependent linear isotropic thermoelastic media.

*5.4. Summary*

The boundary integral equations for temperature, displacement and rotation are summarized as follows:

$$c^{Q*}(\xi)\vartheta(\xi) - \int_S q^{Q*}(\mathbf{x},\xi)\vartheta(\mathbf{x})dS(\mathbf{x}) = -\int_S \vartheta^{Q*}(\mathbf{x},\xi)q(\mathbf{x})dS(\mathbf{x}) + \int_V \vartheta^{Q*}(\mathbf{x},\xi)Q(\mathbf{x})dV(\mathbf{x}) \quad (150)$$

$$c_{ij}^{F*}(\xi)u_i(\xi) + \oint_S t_{ij}^{F*}(\mathbf{x},\xi)u_i(\mathbf{x})dS(\mathbf{x}) + \int_S m_{ij}^{F*}(\mathbf{x},\xi)\omega_i(\mathbf{x})dS(\mathbf{x}) + \int_S h_j^{F*}(x,\xi)\vartheta(x)dS(x)$$
$$= \int_S u_{ij}^{F*}(\mathbf{x},\xi)t_i(\mathbf{x})dS(\mathbf{x}) + \int_S \omega_{ij}^{F*}(\mathbf{x},\xi)m_i(\mathbf{x})dS(\mathbf{x}) + \int_V u_{ij}^{F*}(\mathbf{x},\xi)F_i(\mathbf{x})dV(\mathbf{x}) \quad (151)$$
$$+ \int_S f_j^{F*}(x,\xi)q(x)dS(x) - \int_V f_j^{F*}(\mathbf{x},\xi)Q(\mathbf{x})dV(\mathbf{x})$$

$$c_{ij}^{C*}(\xi)\omega_i(\xi) + \oint_S t_{ij}^{C*}(\mathbf{x},\xi)u_i(\mathbf{x})dS(\mathbf{x}) + \oint_S m_{ij}^{C*}(\mathbf{x},\xi)\omega_i(\mathbf{x})dS(\mathbf{x})$$
$$= \int_S u_{ij}^{C*}(\mathbf{x},\xi)t_i(\mathbf{x})dS(\mathbf{x}) + \int_S \omega_{ij}^{C*}(\mathbf{x},\xi)m_i(\mathbf{x})dS(\mathbf{x}) + \int_V u_{ij}^{C*}(\mathbf{x},\xi)F_i(\mathbf{x})dV(\mathbf{x}) \quad (152)$$

The singular characteristics of the kernel functions are summarized in Tables 1, 2 and 3. These singularities become crucial in the development of the boundary element formulation. Furthermore, we notice that the thermal boundary integral equation is not coupled to the mechanical boundary integral equations, exactly as in the case of classical thermoelasticity.

## 5. Conclusion

The general thermal and mechanical reciprocal theorems have been developed here for the first time for three-dimensional steady state size-dependent thermoelastic response of linear anisotropic solids. For this size-dependent phenomena, a change in temperature can create not only thermal strain deformation, as in the classical case, but also thermal mean curvature deformation can occur. As in steady state classical thermoelasticity, the governing partial differential equations are uncoupled. This means the thermal equation does not depend on mechanical equation. The corresponding boundary integral formulation is based on the thermal and mechanical reciprocal theorems for general anisotropic thermoelastic material. For this purpose, we have also developed the corresponding weak formulations for linear thermoelasticity.

Furthermore, in this paper, the integral equations have been derived for isotropic thermoelastic materials, along with corresponding singular infinite space fundamental solutions or Green's functions. For isotropic materials, the problem simplifies greatly. There is no thermal mean curvature deformation, and the thermoelastic effect is solely the consequence of thermal strain deformation. As a result, the size-dependency for isotropic materials is specified by one characteristic length scale parameter $l$, and the thermal effect is specified completely by the classical thermal expansion coefficient $\alpha$.

Using the present theoretical foundation, numerical implementations of the boundary element formulation for two-dimensional case has been developed [29]. The corresponding axisymmetric and three-dimensional cases will be the subject of forthcoming papers.